\documentclass[preprint]{aastex}

\slugcomment{To appear in The Astrophysical Journal}

\shorttitle{Near-IR spectroscopy of red giants in $\omega$~Centauri
} 
\shortauthors{Origlia et al.}

\begin{document}
 
\title {A near-infrared spectroscopic screening of the 
red giant populations in $\omega$~Centauri}
  
\author{Livia Origlia} 
\affil{INAF - Osservatorio Astronomico di Bologna, Via Ranzani 1,
I-40127 Bologna, Italy}                   
\email{origlia@bo.astro.it}

\author{Francesco R. Ferraro} 
\affil{Dipartimento di Astronomia, Universit\`a di Bologna, 
Via Ranzani 1, I-40127 Bologna, Italy}                   
\email{ferraro@bo.astro.it} 

\author{Michele Bellazzini}
\affil{INAF - Osservatorio Astronomico di Bologna, Via Ranzani 1,
I-40127 Bologna, Italy}                   
\email{bellazzini@bo.astro.it}
 
\author{Elena Pancino}
\affil{Dipartimento di Astronomia, Universit\`a di Bologna, 
Via Ranzani 1, I-40127 Bologna, Italy}                   
\email{pancino@bo.astro.it}

\altaffiltext{1}{Based on observations at the ESO NTT 3.5m telescope 
using the near-infrared imager\-/\-spectrometer SOFI within the observing 
programs 64.N-0038 and 68.D-0287.
}

\begin{abstract}

Near-infrared spectra of 21 giants in $\omega$~Centauri, 
spanning the whole range of metallicities observed in this cluster, 
are presented.  
This work is part of a coordinated
photometric and spectroscopic campaign in the optical and in the infrared,
aimed at studying the complex stellar population of $\omega$~Centauri and 
understanding its formation and chemical evolution.
By analyzing the several CO and OH molecular bands and atomic lines in the
spectra of the selected giants, metal abundances and abundance ratios 
have been obtained.
The existence of three major metallicity regimes
at [Fe/H]$\approx$-1.6, $\approx$-1.2
and [Fe/H]$\le-0.5$ has been confirmed.   
The most metal-rich stars in our sample show a 
lower (if any) $\alpha$-enhancement  
when compared to the more metal-poor components, 
suggesting that they
should have formed in a medium significantly polluted 
by type~Ia supernova ejecta. 
Isotopic carbon abundances have been 
also inferred, providing an average $^{13}$C/$^{12}$C$\simeq$4, 
which clearly indicates that extra-mixing processes occurred 
in the stellar interiors 
during the ascent on the Red Giant Branch.

\end{abstract}
 

\section{Introduction}

$\omega$~Centauri is the most massive globular cluster of our Galaxy 
(Pryor \& Meylan 1993)
with a remarkable flattening ascribed to its high degree of rotation.
It has been the subject of the largest chemical
(see e.g. Norris \& Da Costa 1995, Norris, Freeman \& Mighell 1996, 
Suntzeff \& Kraft 1996)
and kinematical investigations 
(see e.g. Merritt, Meylan \& Mayor 1997, van Leeuwen et al. 2000)
ever attempted on a globular cluster.
The global picture emerging from these surveys is that this
is the only Galactic globular cluster showing an undoubted
spread of metallicity, with a major distribution peak at 
${\rm [Fe/H]}\simeq -1.6$
and a long, extended tail up to ${\rm [Fe/H]}\simeq -0.5$, possibly with a 
second, smaller peak at ${\rm [Fe/H]}\simeq -1.2$.
The more metal-rich stellar component seems also more centrally
concentrated and kinematically cooler than the metal-poor one
(Norris et al. 1997).
Besides, while the latter has a well defined systemic rotation, the
former shows no evidence of it (Norris et al. 1997), 
somewhat in contrast with a simple, dissipative formation scenario.
 
The detailed chemical composition of $\omega $~Centauri giants is
even more complex and puzzling.
The overall over-abundance of $\alpha $-elements with respect to the Sun
indicates a primordial enrichment
by type~II supernovae (SNe) with massive progenitors.
The observed enhancement of s-process elements 
(see e.g. Lloyd Evans 1983, Smith, Cunha \& Lambert 1995, Smith et al. 2000, 
Vanture, Wallerstein \& Suntzeff 2002)               
towards higher metallicities suggests a more extended
period of primordial enrichment (possibly as long as a few Gyrs)
provided by AGB stars of intermediate-low mass.
Concerning the CNO group elements,
while the carbon depletion and nitrogen enhancement
shown by the bulk of metal-poor stars 
(Norris \& Da Costa 1995) can 
be easily explained with evolutionary mixing effects
along the Red Giant Branch (RGB), 
the carbon and nitrogen over-abundances observed in
some CO/CN-strong stars (Bessel \& Norris 1976, Persson et al. 1980)
have been suggested to be of primordial origin 
(Norris \& Da Costa 1995, Hilker \& Richtler 2000), 
indicating once again an interstellar medium pollution by AGB stars. 
In order to explain such over-abundances of s-process and CN 
elements, a self-enrichment scenario can be invoked,
where the products released by intermediate mass AGB stars 
not only pollute the interstellar medium but can be also accreted onto the 
atmospheres of lower mass cluster stars during their Main-Sequence and/or 
Sub-Giant Branch evolution (see e.g. the recent works by Ventura et al. 2001, 
Thoul et al. 2002 and references therein). 
 
A recent burst of wide field photometric studies has finally revealed 
a further, new {\it anomalous} population 
(Lee at al. 1999, Pancino et al. 2000).
In particular, our photometric survey (Pancino et al. 2000) of 
more than 220,000 stars over a $34^{\prime }\times 34^{\prime }$ field of view,
allows to identify a distinct RGB sequence (RGB-a), 
previously unknown, to the red side of the main
cluster sequence.
Six of these {\it anomalous} RGB-a stars are co-identified in the 
Norris et al. (1996)  
spectroscopic sample: all of them are clear cluster members, 
the most metal-rich ones 
having also strong BaII (Lloyd Evans 1983) and CN 
lines (Hilker \& Richtler 2000).
Our photometric analysis indicates that, 
within the complex structure of the $\omega$~Centauri RGB,
three major sub-populations can be distinguished, 
the largest being metal-poor (RGB-MP), 
one moderately populated at intermediate metallicity (RGB-MInt) and   
the {\it anomalous}, metal rich sequence (RGB-a) which 
accounts for $\sim 5 \%$ of the whole stellar content of the cluster.
The RGB-MInt and the RGB-a also show different spatial distributions compared 
to the bulk of metal poor stars (see Pancino et al. 2000 and reference therein). 
 
As part of a large-project (Ferraro, Pancino \& Bellazzini 2002a) 
devoted to study the origin and evolution of the complex stellar population 
in this peculiar cluster, 
we are also undertaking a spectroscopic survey both in the visual 
and in the near-infrared (IR),    
in order to trace the detailed abundance patterns of the $\alpha $, CNO
group and s-process elements of the {\it anomalous},
metal-rich sub-population, as done in the past years for the more 
metal-poor counterparts. 
23 giants in $\omega$~Centauri, 
belonging to both the RGB-a and
the control sample of the {\it normal}, more metal-poor populations,
have been recently observed at high resolution with UVES at VLT 
(Paranal, Chile). 
A first sample of six stars (3 RGB-a and 3 RGB-MInt) 
has been already analyzed (Pancino et al. 2002), providing accurate 
Fe, Cu, Ca and  Si abundances.  
The most metal-rich stars have [Fe/H]$>-0.8$
and are less $\alpha$-enhanced than 
the stars at intermediate metallicity. 
This result is the first evidence that the {\it anomalous}, 
metal-rich RGB-a population 
formed in a medium significantly polluted by type~Ia SNe, 
hence with a substantial different history of chemical enrichment 
compared to the one of the {\it normal} population.  

In this metal-rich regime, the optical spectra of cool giants 
can be severely affected by molecular blending, making the analysis 
more difficult and somewhat uncertain. 
For this reason, a complementary survey in the near IR has been also performed:
we secured a set of medium resolution IR spectra of bright 
giants, spanning the whole range of metallicities of $\omega$~Centauri.
IR spectroscopy is indeed a crucial tool to study the cool
stellar systems, since the IR spectral range is intrinsically 
more sensitive to low temperatures 
compared to the visual one, and also
reddening corrections, blanketing effects and background
contamination by Main-Sequence stars are much less severe, allowing to
properly characterize the red stellar sequences also in 
metal-rich and crowded environments.
The results of our IR spectroscopic campaign are  
presented in this paper.
Sect.~2 presents the observed sample. Sect.~3 describes 
the spectral analysis procedure.
Sect.~4 shows the results of the abundance analysis. 
In Sect.~5 we discuss the-state-of-the-art scenario for the formation and 
chemical enrichment of this peculiar cluster 
and in Sect.~6 we draw our conclusions.

\section{The observed sample}
 
A sample of bright giants in $\omega$~Centauri 
has been observed at the ESO NTT 3.5m telescope (La Silla, Chile) 
using the near-IR imager/spectrometer SOFI 
on January 13th, 2000, and December 27-29, 2001 
as a backup programme in poor seeing conditions. 
Long slit spectra through a 0.6~$arcsec$ slit, with both the
low resolution JH blue grism (R$\simeq$900 at 1.2 $\mu$m) 
and the medium resolution H band grism (R$\simeq$2300 at 1.6 $\mu$m),
have been obtained. 
The spectra have been sky-subtracted, flat-fielded and 
corrected for atmospheric absorption using an 
O-star spectrum as reference.
The wavelength calibration has been computed using 
a Ne-Ar lamp and a cubic spline interpolation. 
The overall signal to noise ratio of the final spectra is $\ge$30.

The observed sample consists of 21 stars belonging to both
the {\it anomalous}, metal-rich (RGB-a) and the {\it normal}, metal-poor
sub-populations (RGB-MInt and RGB-MP) of $\omega$~Centauri 
(according to the nomenclature adopted in Pancino et al. 2000). 
Fig.~1 shows their location  
in the $I_0,(B-I)_0$ color-magnitude diagram, according to our WFI photometry 
and by assuming a reddening E(B-V)=0.12 
(Harris 1996). 
The stellar parameters of the observed stars 
have been established as follows: 
temperature ($T_{\rm eff}$) from optical colors (cf. Fig.~1), by 
using the color-temperature transformation 
of Montegriffo et al. (1998) specifically calibrated on globular cluster giants, 
gravity ($log~g$) from theoretical evolutionary tracks 
according to the location on the RGB, while for the microturbulent velocity 
($\xi$) an average value of 1.5 km/s has been adopted 
(see also Origlia et al. 1997).  
The full sample of observed stars and their adopted photospheric parameters 
are listed in Table~1.
Fig.~2 shows a few examples of the observed infrared spectra. 

\section{Spectral analysis}

The near-IR (1-2.5 $\mu$m) spectra of cool stars  
show many absorption lines due to neutral metals
and molecules (see e.g. Thompson et al. 1969;
Thompson \& Johnson 1974,
Lambert et al. 1984, Wallace \& Hinkle 1996, 
Meyer et al. 1998,  Wallace et al. 2000).\\
Most of the molecular lines in the H band are not saturated and can be safely 
treated under the LTE approximation, being roto-vibrational transitions in the 
ground electronic state, providing accurate C and O abundances. 
The abundance analysis of these elements in the optical is somewhat 
more critical, particularly in the most metal-rich and coolest stars,
where the TiO and CN blends and/or NLTE corrections can severely affect 
for example both the [OI] and OI-triplet lines 
(see e.g. Melendez \& Barbuy 2001, Hill 2001). 
The neutral, atomic lines in the J and H bands are heavily 
saturated (lying on the damping part of the curve of growth) and
somewhat less sensitive to element abundance variations,
with respect to the molecular lines. 
All these lines have relatively high excitation potentials and 
form quite deeply in the stellar atmospheres, where the LTE regime 
still dominates, even in giants with low temperatures and gravities. \\
A reasonable number of these lines are strong, 
and not affected
by severe blending, hence they can be also measured at low-medium resolution
(see e.g. Kleinmann \& Hall 1986; Origlia et al. 1993;
Wallace \& Hinkle 1997; Joyce et al. 1998; Frogel et al. 2001; 
Smith, Terndrup \& Suntzeff 2002),
making them powerful abundance tracers not only in stars but also in more distant 
stellar clusters and galaxies and in a wide range of metallicities and ages
(cf. Origlia et al. 1997; Oliva \& Origlia 1998, Origlia 2000).       

Performing abundance analysis from low-medium resolution spectra is not an easy 
task. It requires a full spectral synthesis technique to account for 
line blending and eventually cross-check calibrations with high resolution 
spectra to select the most sensitive and less contaminated lines. 
Recently, the latter task has became possible thanks to 
the advent of the first generation of IR echelle spectrographs with high resolution. 
Comparison between the overall metallicities
as well as abundance patterns inferred from high resolution IR spectra of 
bright giants in Bulge globular clusters (Origlia, Rich \& Castro 2002) 
with those obtained from medium resolution spectra of the cluster core   
(Origlia et al. 1997, Origlia 2003) 
shows a full consistency of the two sets of measurements 
well within $\pm$0.2 dex. 
 
Synthetic spectra of giant stars 
for different input atmospheric parameters 
have been computed, using an updated
version of the code described in Origlia et al. (1993)
(see also Origlia et al. 2002). 
The code uses the LTE approximation and is based
on the molecular blanketed model atmospheres of
Johnson, Bernat \& Krupp (1980)
which have been extensively used for abundance
studies based on high resolution spectra of cool stars 
(e.g. Lambert et al. 1984, Smith \& Lambert 1985, 1990).
For temperatures higher than 4000~K the ATLAS9 models are used.
Since in the IR the major source of continuum opacity is H$^-$  
with its minimum near 1.6 $\mu$m, 
the dependence of the results on the choice of model
atmosphere is much less critical than in the visual range.                                          
Three main compilations of 
atomic oscillator strengths are used, namely 
the Kurucz's database, Bi\`emont \& Grevesse (1973, hereafter BG73)
and Mel\'endez \& Barbuy (1999, hereafter MB99).
On average, MB99 log-gf values are systematically lower than Kurucz ones but
in most cases the difference does not exceed 0.2~dex 
and the overall scatter in the derived abundances is $<0.1$~dex 
(see also Origlia et al. 2002  
for a more quantitative comparison). 
Voigt profiles (see e.g. Mihalas 1970) 
have been computed by taking into account thermal, natural and Van der Waals 
broadenings (see e.g. Gray 1999).
The dissociation energies of the molecules are from Table~2 in 
Johnson et al. (1980), the reference Solar abundances from  
Grevesse \& Sauval (1998).    
 
Our code provides full spectral synthesis over the 1-2.5 $\mu$m range, 
allowing to obtain abundance estimates both by best-fitting the full 
observed spectrum (cf. Fig.~2) and by measuring the equivalent widths 
of selected lines. 
The equivalent widths have been measured by performing a Gaussian fit 
to the line profile, typical values ranging between 100 and 1000 m\AA, 
with a conservative error of $\pm$150~m\AA\  
which implies an average uncertainty in the resulting 
abundances of $\approx$0.2-0.3~dex. 

The medium resolution H-band spectra also allow us to estimate radial 
velocities with an accuracy of $\pm$20 km/s.
All the observed stars can be reasonably considered cluster members, 
having heliocentric radial velocities of $\approx$240~km/s, 
consistent with the 
measured systemic velocity of the cluster  
(Pryor \& Meylan 1993, Harris 1996).
Six giants in our sample 
(WFI217134=ROA167, WFI221132=ROA300, WFI246325=ROA513,  
WFI225201\-=\-ROA8141, WFI222741=ROA8161 and WFI236826=ROA9008)   
have accurate radial velocity measurements 
in the range 220-257 km/s (Mayor et al. 1997), while all of them 
have also an identified counterpart in 
the catalog of proper motions by van Leeuwen et al. (2000). 
Their cluster membership probability, according to van Leeuwen et al. (2000), 
is always $\ge$99\% with the exception of WFI237727 whose value is 41\%.

\subsection{Error budget}

Three main sources of uncertainty can affect the derived elemental abundances, 
namely the errors in the equivalent width 
measurements, the uncertainty in the stellar parameter assumptions and 
in the overall fitting procedure.

In order to better quantify the impact of the errors in the equivalent 
width measurements   on the inferred abundances we computed the curves of 
growth for the neutral atoms.  
As an example, Fig.~3 shows the curves of growth  
of FeI computed from our synthetic spectra in the J band.
In our stellar spectra three major FeI lines at 
$\lambda$1.1783, $\lambda$1.1882 and $\lambda$1.1973 $\mu$m 
are observed (see Fig.~2) 
and the measured equivalent widths well overlap the theoretical curve, 
as shown in Fig.~3. 
According to the curves of growth,
an uncertainty of $\pm$150~m\AA\ in the equivalent widths of the 
two brightest lines at $\lambda$1.1882 and $\lambda$1.1973 $\mu$m will  
imply an error in the inferred abundance of $\le$0.2 dex in the high 
metallicity regime and $\approx$0.3 dex at low metallicity.
The iron abundance inferred from the fainter FeI line at $\lambda$1.1783 
$\mu$m is more uncertain.  

Different assumptions in the input stellar parameters can also influence 
the derived abundances. For the low-mass giants under consideration,  
$\Delta $T$_{\rm eff}$$\pm$200~K, $\Delta $log~g=$\pm$0.5 and
$\Delta \xi=\pm$0.5~~km~s$^{-1}$ are somewhat conservative estimates
of the possible uncertainties.
By varying the stellar parameters within the above ranges,
the change in the inferred element abundances is always $\le$0.2 dex 
(see also Origlia et al. 2002 for a more 
exhaustive discussion). 
Moreover, the simultaneous synthesis of all the atomic and molecular lines, 
which have a different sensitivity to the various stellar parameters, 
puts more stringent constraints on the best-fit solution, making 
the latter more robust.             
As shown in Fig.~4 (top panels), models with $\pm$0.2~dex 
variation of the elemental abundances with respect to the best-fit solution 
generate quite different line profiles, even at low-medium resolution, 
particularly in the case of the molecular lines which are not saturated 
and hence more sensitive to the element abundance. 

While fitting observed spectra with synthetic models,
it can be also useful to quantify the goodness of the 
fits in a statistical way.  
The adopted process of best fitting is performed by varying the elemental 
abundances in the synthetic spectrum until the best match with the
observed features is found. 
Synthetic spectra with slightly lower elemental abundances should be 
{\em systematically} shallower than the best-fit
solution, while the opposite occurs
when higher abundances are adopted.       
To take full advantage of these {\em systematic} differences, 
as a statistical function of merit we adopt 
the difference between the model and the observed spectrum 
($\delta_{mod-obs}$, hereafter $\delta$, for brevity). 
When the best-fit solution is adopted, $\overline{\delta}$ 
should be $\approx 0.0$. 
Synthetic spectra with lower
elemental abundances with respect to the best-fit 
solution will give $\overline{\delta} < 0.0$, while  
when higher abundances are adopted $\overline{\delta} > 0.0$. 
In order to quantify systematic discrepancies, this parameter is 
more powerful than the classical $\chi ^2$ test, which is instead 
equally sensitive to {\em random} and {\em systematic} scatters.  

Since $\delta$ is expected to follow a Gaussian distribution,
we compute $\overline{\delta}$ and the corresponding standard deviation
($\sigma$) for the best-fit solution and 6 {\it test models} 
with abundance variations $\Delta[X/H]=\pm$0.2, 0.3, 0.4 dex  
with respect to the best-fit.
We then extract 10000 random subsamples from each 
{\it test model} (assuming a Gaussian distribution) 
and we compute the probability $P$ 
that a random realization of the data-points around
a {\it test model} display a $\overline{\delta}$ that is compatible 
with the {\em best-fit} model.                          
$P\simeq 1$ indicates that the model is a good representation of the 
observed spectrum.

As an example of the application of such a statistical test,  
the lower panels of Fig.~4 show the results for the three 
portions of the WFI246325 stellar spectrum 
shown in the corresponding upper panels. 
It can be easily appreciated that the best-fit solution provides 
in all cases a clear maximum in $P$ ($>$99\%) 
with respect to the {\it test models}. 
More relevant, {\it test models} with an abundance 
variation $\Delta[X/H]\ge\pm 0.3$ dex 
lie always at $\ge$2 $\sigma$ from the best-fit solution,
demonstrating that our spectral analysis allows to fully appreciate 
elemental abundance variations of $\pm0.2-0.3$ dex.       

\section{Results}
 
With the present infrared spectroscopic survey we have been able to 
measure the abundances of key metals like iron, 
carbon, oxygen and other $\alpha $-elements 
in the full range of metallicities (see Table~1).  
Carbon and oxygen abundances are inferred from  
the CO ($\Delta v=3$) and OH ($\Delta v=2$) molecular lines in
the H (1.5-1.8 $\mu$m) band, while  
Fe, Mg, Si and Ca abundances are derived from a few major neutral lines 
both in the J and H bands.
These elements play a crucial role in disentangling
evolutionary mixing effects and primordial enrichment, a fundamental
step to answer some more general - and still pending - questions
regarding the global picture of the formation and evolution
of this enigmatic globular cluster. 

The abundance analysis on the giants in our sample confirms 
the existence of three major metallicity 
regimes in $\omega$~Centauri, namely:
metal-poor (RGB-MP, 7 giants) at [Fe/H]=$-1.6\pm 0.2$ dex, 
intermediate (RGB-MInt, 7 giants) at [Fe/H]=$-1.2\pm$0.2~dex 
and metal-rich (RGB-a, 7 giants) at $-0.9\le$[Fe/H]$\le-0.5$.   
According to its location in the $I_0,(B-I)_0$ color-magnitude diagram, 
the intermediate metallicity WFI228136 star (cf. Table~1) 
could likely belong to the Asymptotic Giant Branch. 

Fig.~5 shows the inferred O, Ca, Si, and Mg abundances relative to iron as 
a function of [Fe/H] 
for the observed 21 giant stars in $\omega$~Centauri (see also Table~1). 
All these $\alpha$-elements behave in a similar fashion, confirming 
their common origin as by-products from type~II SN explosions.  
Only the metal poor and intermediate metallicity giants 
are significantly $\alpha $-enhanced, 
while the {\it anomalous},
metal rich component shows almost no enhancement, 
suggesting a major enrichment by type~Ia SNe as well. 
Fig.~6 shows the cumulative [$\alpha$/Fe] {\it versus} [Fe/H] diagram, 
as obtained by averaging 
the single [Ca/Fe], [Mg/Fe], [Si/Fe] and [O/Fe] measurements.  
The RGB-MP and RGB-MInt stars show   
an overall constant [$\alpha $/Fe]$=+0.30\pm0.03$~dex,  
while the most metal-rich RGB-a is progressively    
less $\alpha$-enhanced down to solar [$\alpha $/Fe].

This differential behavior of [$\alpha $/Fe] with metallicity fully confirms 
(on a sounder statistical ground) the result 
of Pancino et al. (2002), based on the  
analysis of a preliminary sample of six giants observed with UVES.
Actually, the three metal-rich stars 
(namely WFI221132=ROA300, WFI222068 and WFI222679) 
in Pancino et al. (2002) sample have been also observed 
in our IR survey and there is full consistency (within $\pm$0.2~dex) 
between the two sets of [Fe/H], [Ca/Fe] and [Si/Fe] abundances and abundance ratios. 
Other six giants in our sample are in common with the sample of  
Norris et al. (1996), who provided calcium abundances by means of low-medium 
resolution spectroscopy in the optical. 
Four of them (WFI225201=ROA8141, WFI222741=ROA8161, WFI236826\-=\-ROA9008 
and WFI217134=ROA167) belong to the metal-poor population and our [Ca/H] abundances 
are systematically slightly larger (on average $\simeq $0.2~dex) than their value, 
but well within the observed spread in this metallicity regime 
(see for example Fig.~8 of Norris et al. 1996, where their 
values are compared with those obtained in previous works by means 
of high resolution spectroscopy). 
Our [Ca/H] abundances for the two other giants 
(WFI221132=ROA300 and WFI246325=ROA513) 
in common with Norris et al. (1996) and belonging to the 
metal-rich population are much lower (about 0.7~dex) than the values reported 
by Norris et al. (1996) but in good agreement with those 
obtained by Pancino et al. (2002).
Very recently, for WFI221132=ROA300 Vanture et al. (2002) 
have obtained even lower values.  
Nevertheless, Norris et al. (1996) themselves mention the fact that 
their largest [Ca/H] values should be treated with caution since they have been 
inferred by extrapolating the calibration at lower metallicities.
These two stars have been also previously classified as S-stars by Lloyd Evans 
(1983), showing strong Ba~II absorption lines.

Fig.~7 shows our inferred [C/Fe] and 
$^{13}$C/$^{12}$C abundance ratios, 
based on the measured CO band heads in the H band,
as a function of the stellar metallicity for the giant stars 
in our sample (cf. Table~1). 
A least square fitting to the observed values gives average [C/Fe]$=-0.20\pm 0.14$ 
and $^{13}$C/$^{12}$C$=3.8\pm 1.0$, without any obvious dependence - within the 
observed scatter - on stellar metallicity or luminosity.  
Our values are fully consistent with $^{13}$C/$^{12}$C$\approx$4-6, 
obtained from high resolution optical spectroscopy 
using the CN lines (Brown \& Wallerstein 1993; Zucker, Wallerstein \& Brown 1996; 
Wallerstein \& Gonzalez 1996; Vanture, Wallerstein \& Suntzeff 2002).    
Very recently, Smith et al. (2002) also obtained medium resolution IR 
spectra of 11 giants with low-intermediate metallicity.
By measuring the CO lines in the K band they found 
[C/Fe] between $\approx$-1.0 and $\approx$0 and $^{13}$C/$^{12}$C$\approx$3-7
without any trend with iron abundance, in perfect agreement with our analysis.

In a pioneering work, Persson et al. (1980) measured the CO photometric 
index at 2.3 $\mu$m of 82 stars in the upper RGB of $\omega $~Centauri, 
spanning a wide range of metallicities.
They found that about 50\% of the stars have an anomalously large CO index 
(CO-strong stars) when compared to the values 
observed in other Population II giants with similar infrared colors 
(CO-weak stars).
Both mixing and/or primordial scenarios have been invoked to explain this  
bimodal distribution, but both show some inconsistency. 
A few years later, for most of the stars 
in the Persson et al. (1980) sample,
Cohen \& Bell (1986) 
derived carbon abundances from the CH bands in the 3000-5000 \AA\ range
at low-medium resolution, without obtaining
any obvious separation between 
the CO-strong and CO-weak groups of Persson et al. (1980).
Norris \& Da Costa (1995) also derived carbon abundances from CH bands 
for a sub-sample of the giants observed by Persson et al. (1980)
using high resolution spectroscopy, and found a somewhat better correlation 
within the CO-strong / CO-weak classification.
Two RGB-a stars (namely ROA300 and ROA513) 
in our sample have been previously observed 
by Persson et al. (1980) and Cohen \& Bell (1986) and can be classified 
as CO-weak stars.\\ 
In this respect, it is worthwhile mentioning that the CO  photometric 
index quickly saturate with decreasing stellar temperature and gravity, 
even at relatively low metallicities ($\sim$1/10 solar), becoming poorly 
sensitive to carbon abundance and 
mainly sensitive to microturbulent velocity variations 
(see e.g. Lambert et al. 1984, Origlia et al. 1993; 1997). 
In the range of stellar parameters covered by the observed giants of 
$\omega$~Centauri, a few hundredths of magnitude 
variation in the CO photometric 
index can be equally accounted for by $\pm$0.3~dex variation in the 
carbon abundance or $\pm$1~km/s 
variation in microturbulent velocity. 
Hence, the spread in the CO index among the $\omega$~Centauri giants 
should not be necessarily/univocally interpreted as a 
carbon abundance anomaly.

\section{Discussion}    

A detailed screening of the abundance patterns in a stellar system is 
crucial to understand the history of its formation and chemical enrichment.
Different elements can be released in the interstellar medium by stars 
with different mass progenitors and in different evolutionary stages, 
thus enriching the gas on different timescales.
Moreover, different elements can undergo different nucleosynthesis processing 
in the stellar interiors, thus providing crucial constraints on the so-called 
{\em evolutionary} versus {\em primordial} scenarios.

\subsection{[$\alpha$/Fe] abundance ratios and primordial enrichment}

$\alpha -$elements are
predominantly released by type~II SNe,
while iron--like elements are mainly produced by 
type~Ia SNe with intermediate mass progenitors. 
Type~Ia SNe are believed to release their processed material only after
$\sim 1 ~Gyr$ from the local onset of star formation (the exact timescale 
depending on the actual star formation history, see e.g. Matteucci \& Recchi 2001 
for a more exhaustive discussion), while enrichment by
type~II SNe occurs virtually from the very beginning.
It is currently accepted that $\alpha $-elements (at least the most abundant isotopes) 
as measured in the stellar photospheres, mainly trace the initial chemical structure of 
the gas from which the stars formed, even though some evolutionary 
effects cannot be excluded (see e.g. Kraft 1994).
In this scenario, the amount of $[\alpha/Fe]$ enhancement compared to the solar values 
and its behavior with the overall stellar metallicity are fundamental tools to constrain 
the star formation and chemical enrichment history 
of a stellar system (see e.g. McWilliam 1997). 

The over-abundance of $\alpha $ 
elements observed in the low-intermediate 
metallicity populations of $\omega $~Centauri 
(see also e.g. Lloyd Evans 1983, Smith, Cunha \& Lambert 1995, Smith et al. 2000) 
suggests that no significant iron enrichment by type~Ia SNe occurred,
either because the star formation process ended before the onset of type~Ia SNe,
or because type~Ia SN winds were efficient in removing most of their 
own processed material (Smith et al. 2000).
Type~I SNe should indeed explode in a lower gas density medium 
compared to type~II SNe and their winds can be more 
efficient in removing their processed material 
(see e.g. Recchi, Matteucci \& D'Ercole 2001), somewhat delaying 
or slowing down the overall iron enrichment of the interstellar medium.

The evidence of a significant decrease of the [$\alpha $/Fe]
over-abundance in the most metal-rich population 
of $\omega $~Centauri, as shown in this paper (see also Pancino et al. 2002), 
conversely suggests 
that the gas from which it formed should have been 
enriched in iron by type~Ia SNe. 
Within the simple picture of pure self-enrichment, 
this would also imply some age spread ($\ga 1$~Gyr) among  
the different populations, as 
indeed suggested by some authors (Hilker \& Richtler 2000, 
Hughes \& Wallerstein 2000). 
Different kinematical (radial velocities -- Norris et al. 1997; 
rotation -- Merrit et al. 1997; 
proper motions -- Ferraro et al. 2002b) and structural 
(ellipticity, Pancino et al. 2000) properties 
among the various sub-populations have been also found.
   
\subsection{$^{12}C/^{13}C$ abundance ratio and stellar mixing}

During the ascent along the RGB, stars undergo a deepening of the 
convective envelope, dredging up (the so-called {\it first dredge-up} 
mixing process) the CN-processed material 
and mixing it with the one in the outer layers of the stellar  
atmospheres (see e.g. Charbonnel 1994 and reference therein).
In low-mass stars the {\it first dredge-up} should occur at the base of the RGB
and, as a consequence of such a mixing, models predict a decrease of  
$^{12}$C and an increase of  
$^{13}$C and $^{14}$N abundances, by amounts which mainly depend on the chemical 
composition and the extent of the convective zone.
However, many sets of observations (see e.g. Suntzeff \& Smith 1991, Shetrone 1996, 
Gratton et al. 2000 and references therein) on low mass field and cluster giants 
provide quite different isotopic abundance ratios compared to those predicted by 
standard models. In particular, much lower $^{13}$C/$^{12}$C ratios compared 
to those expected from the {\it first dredge-up} are inferred, 
suggesting that some other non-standard mixing mechanisms should be at work 
during the ascent of the RGB (see e.g. Charbonnel 1995, Denissenkov \& Weiss 1996, 
Cavallo, Sweigart \& Bell 1998 and references therein).
As a consequence of these additional mixing mechanisms, 
the material from the convective envelope is first brought down to the hotter region 
where it undergoes further nuclear processing 
(the so-called {\it cool bottom processing}), and then it is transported 
back up to the convective envelope. 
An accurate abundance determination of the CNO-group elements is then crucial to 
constrain the mixing processes in the stellar interiors. 

Boothroyd \& Sackmann (1999) 
computed CNO isotopic abundances for different stellar masses and metallicities 
after the {first dredge-up} and the     
{\it cool bottom processing}. 
In the range of metallicities covered by our sample of bright 
(L/L$_{\odot}>$100) giants in $\omega$~Centauri, 
the {\it first dredge-up} mixing models for a solar mass star predict a modest 
($\simeq -0.1$~dex) 
[C/Fe] depletion and a $^{13}$C/$^{12}$C ratio of $\simeq $40.
The {\it cool bottom processing} models predict a further decrease of the 
[C/Fe] and $^{13}$C/$^{12}$C relative abundances, 
with a somewhat larger depletion at lower metallicities.
However, as shown in Fig.~7, in the range of metallicities 
under consideration, 
the [C/Fe] ratio is poorly suitable to trace possible extra-mixing effects, 
being the overall carbon depletion within a factor of 2.
On the contrary,
the $^{13}$C/$^{12}$C ratio is extremely sensitive,  
being further depleted by one order of magnitude 
after the {\it cool bottom processing}. 

All the spectroscopic measurements of the isotopic carbon abundances in 
$\omega$~Centauri have shown very low $^{13}$C/$^{12}$C ratios ($<$10), 
without any obvious trend with metallicity.
A comparison with Boothroyd \& Sackmann (1999) mixing models
suggests that such values cannot be explained by the canonical 
{\it first dredge-up} mixing, but further {\it cool bottom processing} is required. 
In the past years somewhat different algorithms and/or physical scenarios 
have been proposed 
to compute {\it cool bottom processing} models 
(see Weiss, Denissenkov \& Charbonnel 2000 for a review). 
Although a solid physical picture describing the deep mixing processes  
and the detailed evolution of the different abundance patterns is 
still lacking and matter of debate,   
all models agree on the fact that the low 
$^{13}$C/$^{12}$C values as observed in the bright giants 
of $\omega$~Centauri are undoubted signatures of extra-mixing 
occurred in the stellar interiors during the evolution along the RGB.

\section{Conclusions}    

A spectroscopic screening of the multi-populations of red giants in 
$\omega $~Centauri has been performed by analyzing medium resolution 
infrared spectra of 21 bright stars. The major results can be summarized as follow.\\
{\it i)}~
The existence of three major metallicity regimes, namely
metal-poor at [Fe/H]=$-1.6\pm$0.2 dex,
intermediate at [Fe/H]=$-1.2\pm$0.2~dex
and metal-rich at [Fe/H]$\le-0.5$ has been confirmed.\\   
{\it ii)}~
The metal-poor and intermediate metallicity populations show  
$\alpha $-enhancement, tracing a major enrichment of the interstellar 
medium by type~II supernovae on a relatively short time-scale, before the  
onset of type~Ia supernovae.\\
{\it iii)}~
The {\it anomalous}, metal-rich population is significantly less $\alpha $-enhanced 
than the main body of the cluster, 
indicating a major iron pollution of the interstellar medium by type~Ia supernova 
ejecta on longer timescales.\\
{\it iv)}~
The overall metallicity spread and the different degree of $\alpha$-enhancement 
between the {\it normal}, metal poor and 
the {\it anomalous}, metal-rich populations suggests a rather complex chemical 
evolution history, possibly characterized by self-enrichment and implying  
a more continuum star formation process as well as some age spread. \\
{\it v)}~
Isotopic carbon abundances have been measured: 
an average $^{13}$C/$^{12}$C$\simeq$4 has been inferred, providing a univocal 
signature of extra-mixing processes occurred in the interiors of the 
observed bright giants. \\

\acknowledgments
The financial support by the Agenzia Spa\-zia\-le Ita\-lia\-na (ASI) and 
the Ministero dell'Istru\-zio\-ne, Universit\`a e Ricerca (MIUR) is 
kindly acknowledged.

\clearpage

\begin{deluxetable}{llccccccccc}
\tablewidth{18.2cm}
\tablecaption{Observed stars, adopted stellar parameters and inferred abundances.}
\tablehead{
\colhead{star$^a$}&
\colhead{ROA$^b$}&
\colhead{T$_{eff}$}&
\colhead{log~g}&
\colhead{[Fe/H]}&
\colhead{[O/Fe]}&
\colhead{[Ca/Fe]}&
\colhead{[Mg/Fe]}&
\colhead{[Si/Fe]}&
\colhead{[C/Fe]}&
\colhead{$^{12}$C/$^{13}$C}
}
\startdata
\hline
RGB-a&&&&&&&&&&\\
\hline
222068&        &4250 &1.5 & --0.55  &--0.10 &--0.05 &--0.02 &+0.05  &--0.43 & 3.2 \\
227309&        &5000 &2.0 & --0.55  &--0.05 &+0.05  &+0.02  &--0.05 &--0.03 & 3.2 \\   
222679&        &4250 &1.5 & --0.60  &+0.00  &-0.10  &+0.05  &+0.00  &--0.30 & 4.0 \\
226951&        &4250 &1.5 & --0.70  &+0.13  &+0.00  &+0.12  &+0.10  &--0.08 & 3.2 \\
221132& 300    &4000 &1.0 & --0.77  &+0.10  &--0.03 &+0.11  &+0.12  &--0.43 & 3.6 \\
224245&        &4250 &1.5 & --0.80  &+0.13  &+0.10  &+0.10  &+0.20  &+0.00  & 4.0 \\
246325& 513    &4000 &1.0 & --0.90  &+0.15  &+0.20  &+0.17  &+0.20  &--0.20 & 4.0 \\
\hline
RGB-MInt&&&&&&&&&&\\
\hline
225687&        &5000 &2.0 & --1.00  &+0.30 &+0.30 &+0.25 &+0.20 &--0.22 & 5.0 \\   
232173&        &4250 &1.0 & --1.10  &+0.20 &+0.30 &+0.30 &+0.20 &--0.38 & 3.2 \\
229257&        &4500 &1.5 & --1.10  &+0.30 &+0.20 &+0.40 &+0.20 &--0.28 & 3.2 \\   
228136$^c$&    &5000 &2.0 & --1.15  &+0.35 &+0.45 &+0.25 &+0.45 &--0.07 & 5.0 \\   
236094&        &4250 &1.0 & --1.20  &+0.25 &+0.20 &+0.30 &+0.30 &--0.25 & 2.5 \\
224886&        &4500 &1.5 & --1.20  &+0.30 &+0.20 &+0.35 &+0.30 &--0.25 & 2.5 \\   
230340&        &4250 &1.0 & --1.23  &+0.33 &+0.43 &+0.23 &+0.23 &--0.29 & 5.0 \\
\hline
RGB-MP&&&&&&&&&&\\
\hline
217134&167     &4250 &1.0 & --1.53  &+0.33 &+0.43 &+0.23 &+0.23 &--0.09 & 5.0 \\   
236826&9008    &4500 &1.5 & --1.55  &+0.32 &+0.45 &+0.25 &+0.25 &--0.07 & 5.0 \\
237727&        &4250 &1.0 & --1.55  &+0.35 &+0.35 &+0.27 &+0.35 &--0.23 & 3.2 \\
222741&8161    &4500 &1.5 & --1.57  &+0.27 &+0.47 &+0.33 &+0.17 &--0.05 & 5.0 \\
225201&8141    &4500 &1.5 & --1.60  &+0.37 &+0.40 &+0.10 &+0.30 &--0.25 & 2.5 \\
239137&        &4500 &1.5 & --1.60  &+0.30 &+0.30 &+0.30 &+0.40 &--0.38 & 3.2 \\
235385&        &4250 &1.0 & --1.65  &+0.45 &+0.35 &+0.25 &+0.30 &+0.03 & 5.0 \\
\enddata
\tablecomments{
$ $\\
$^a$ WFI catalog number from Pancino et al. (2000).\\
$^b$ Royal Astronomical Observatory - ROA- number from Woolley (1966).\\
$^c$ Possible AGB star candidate.}
\end{deluxetable}

\clearpage

\begin{figure}
\epsscale{1.0}
\plotone{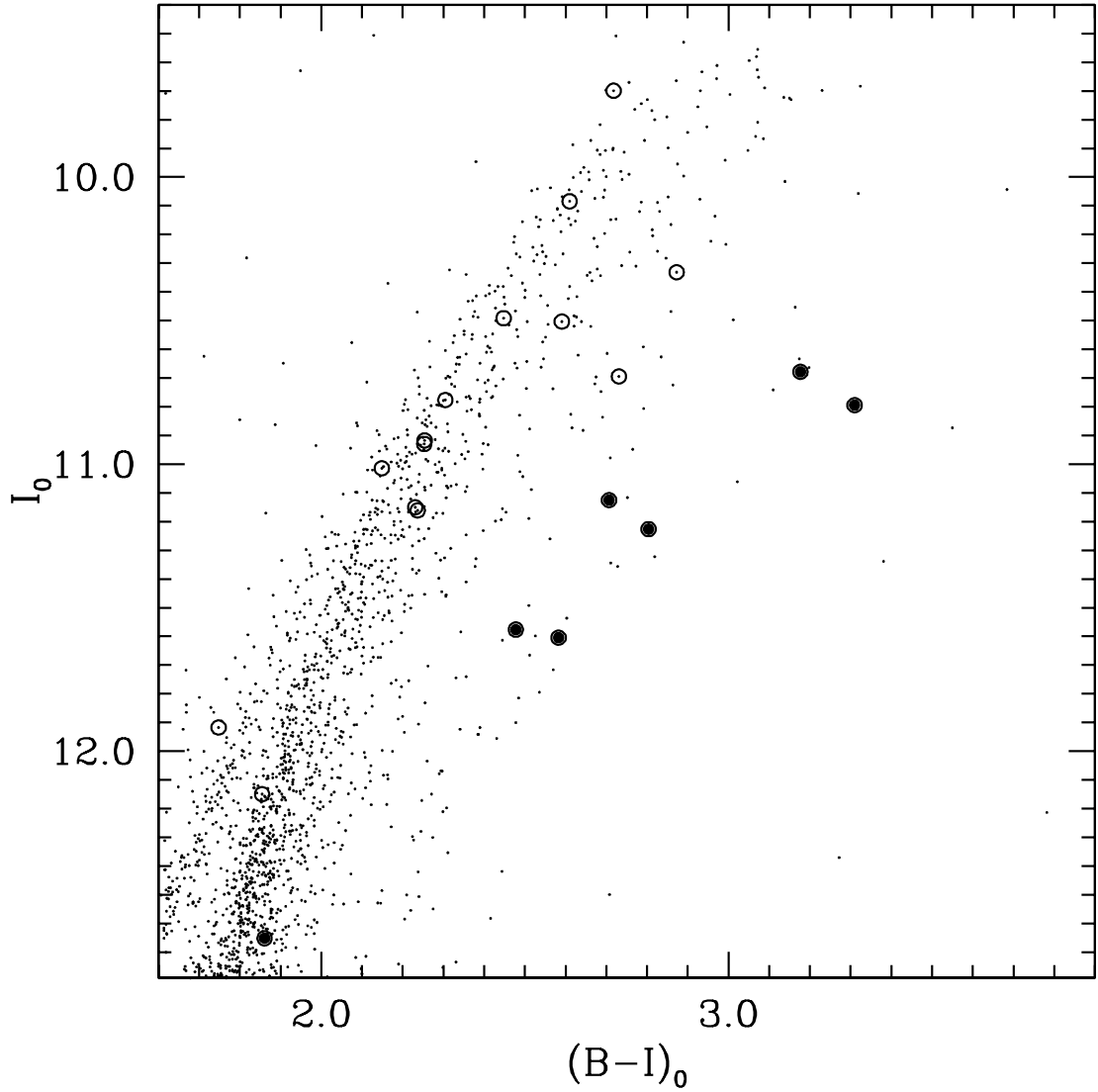}      
\caption{
Location in the $I_0,(B-I)_0$ color-magnitude diagram 
of the observed {\it normal} (open circles) and {\it anomalous} 
(filled circles) populations of $\omega$~Centauri giants 
(cf. Sect~2). 
}
\end{figure}  

\begin{figure}
\epsscale{1.0}
\plotone{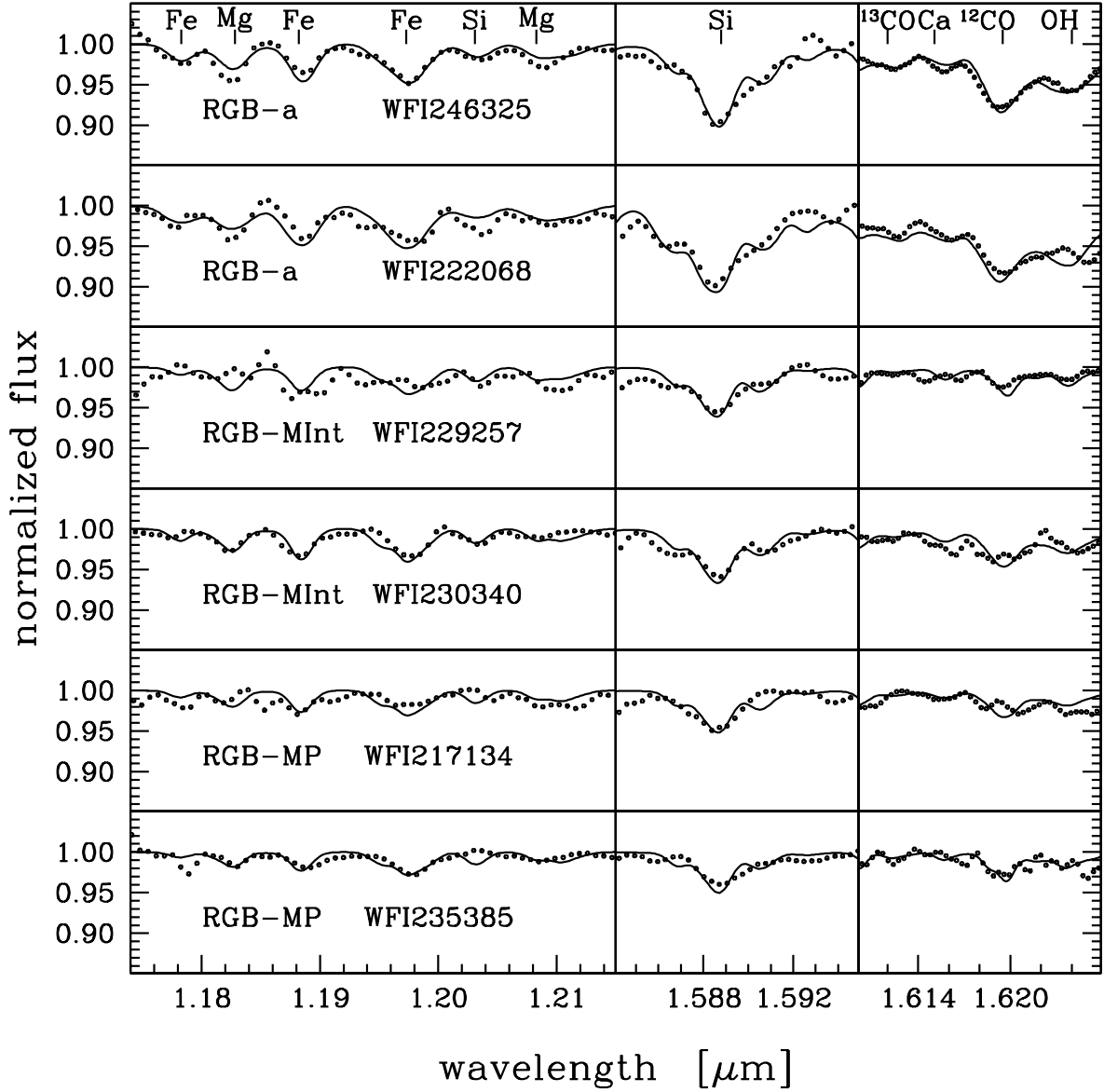}      
\caption{
Normalized long slit spectra (circles)
and our best-fit solutions (full lines) 
of some observed giants, representative of the three metallicity regimes 
inferred in $\omega$~Centauri:
metal-poor (RGB-MP), intermediate metallicity (RGB-MInt) and the {\it anomalous}, 
metal-rich (RGB-a) populations 
(cf. Sect.~2 and Table~1). 
}
\end{figure}  

\begin{figure}
\epsscale{1.0}
\plotone{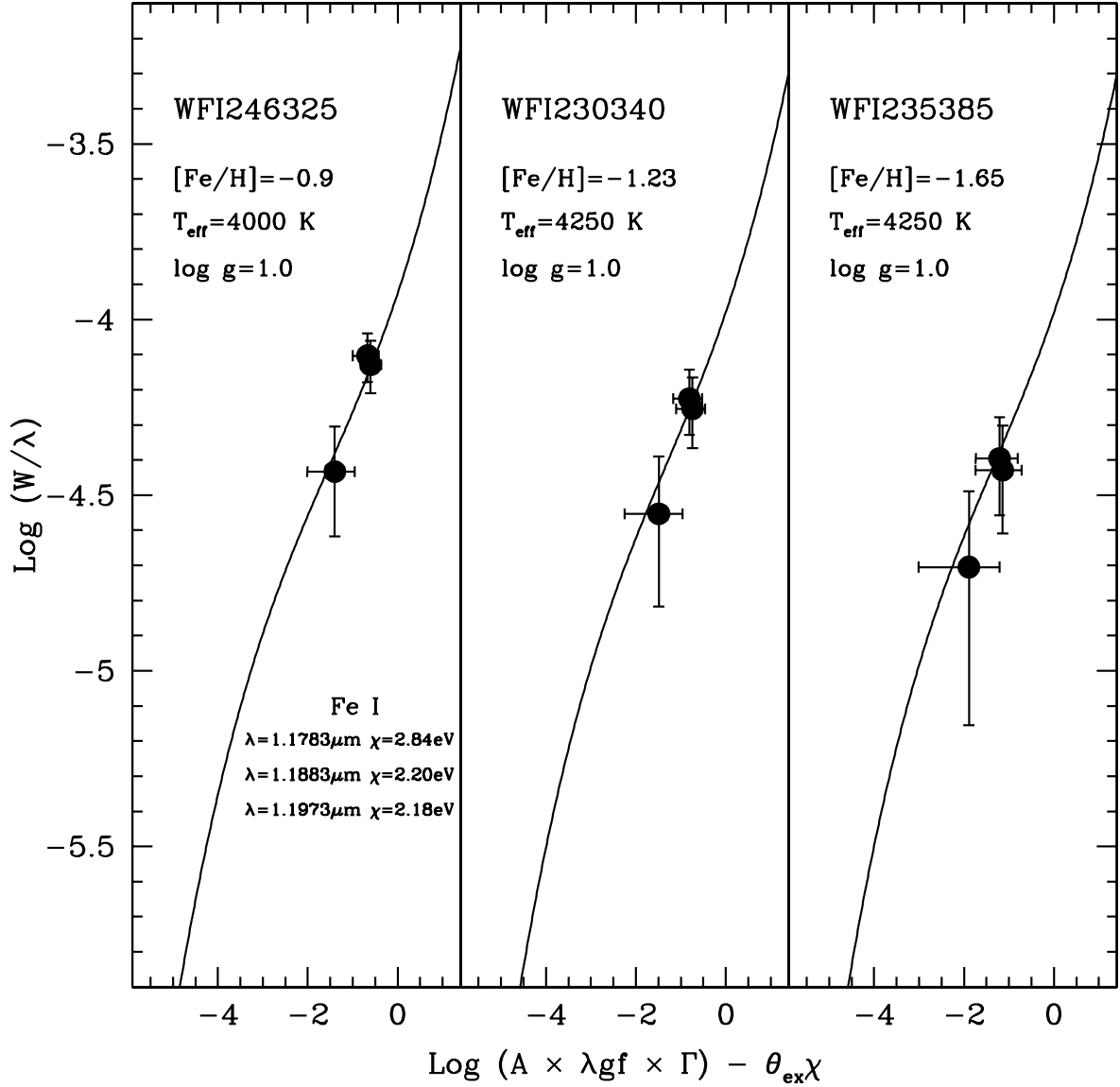}      
\caption{
Curves of growth of FeI computed from synthetic IR spectra in the J band (full lines).
The values measured in 3 giants of our sample, representative of the three 
metallicity regimes (see Sect.~4 and Table~1) are overplotted (full dots). 
The adopted stellar parameters together with  the wavelengths and 
excitation potentials of the FeI lines measured in our observed spectra 
are also indicated.
}
\end{figure}  

\begin{figure}
\epsscale{1.0}
\plotone{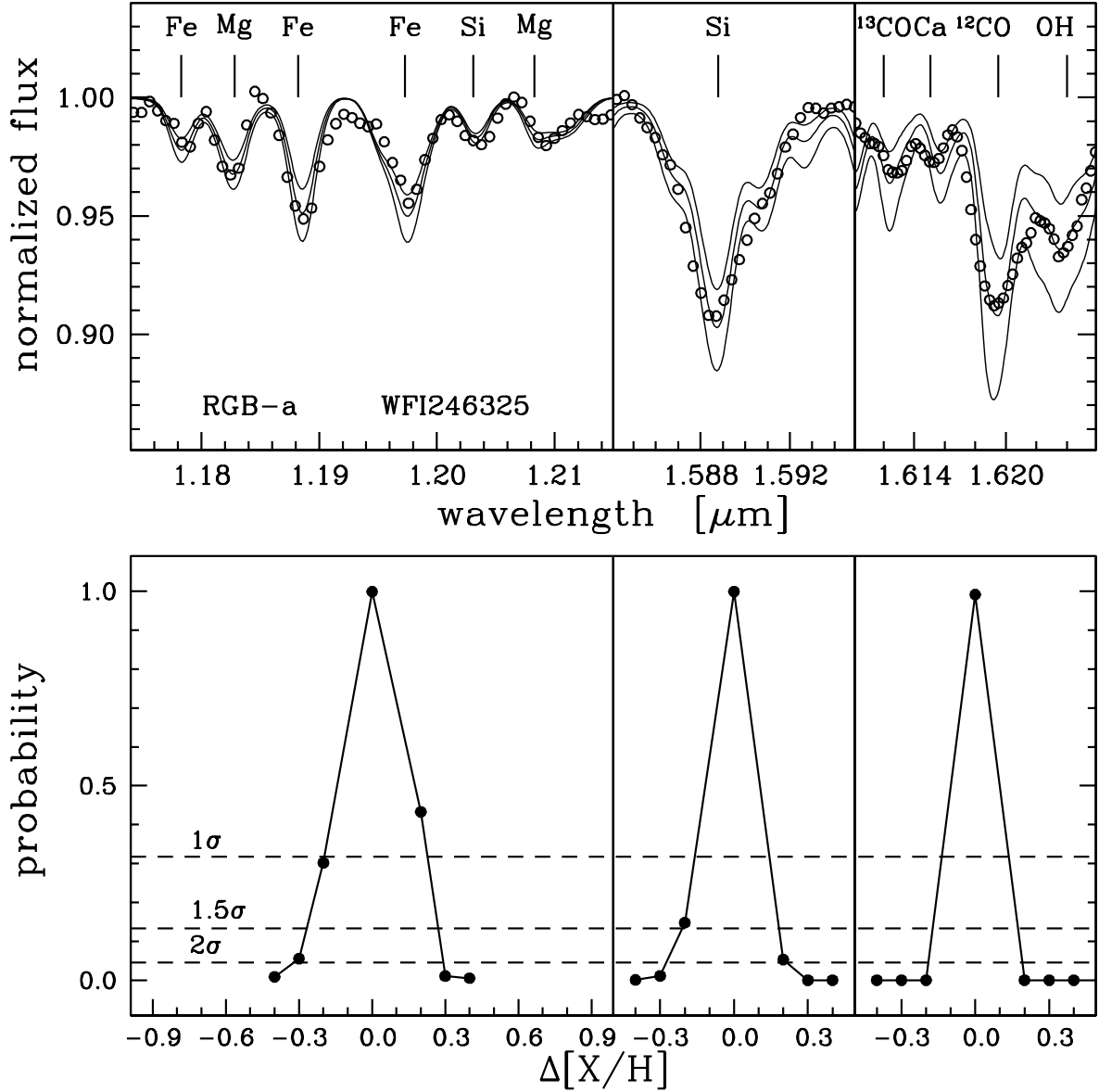}      
\caption{
Top panels:
normalized long slit spectra (circles) of WFI~246325,
our best-fit model and two additional models with $\pm$0.2~dex variation 
of the element abundances with respect to the best-fit solution (full lines).
Bottom panels: probability of a random realization 
of our best-fit solution with varying the elemental abundances 
$\Delta[X/H]$ of $\pm$0.2, 0.3 and 0.4 dex with respect to 
the best-fit (see Sect.~3.1).}
\end{figure}  

\begin{figure}
\epsscale{1.0}
\plotone{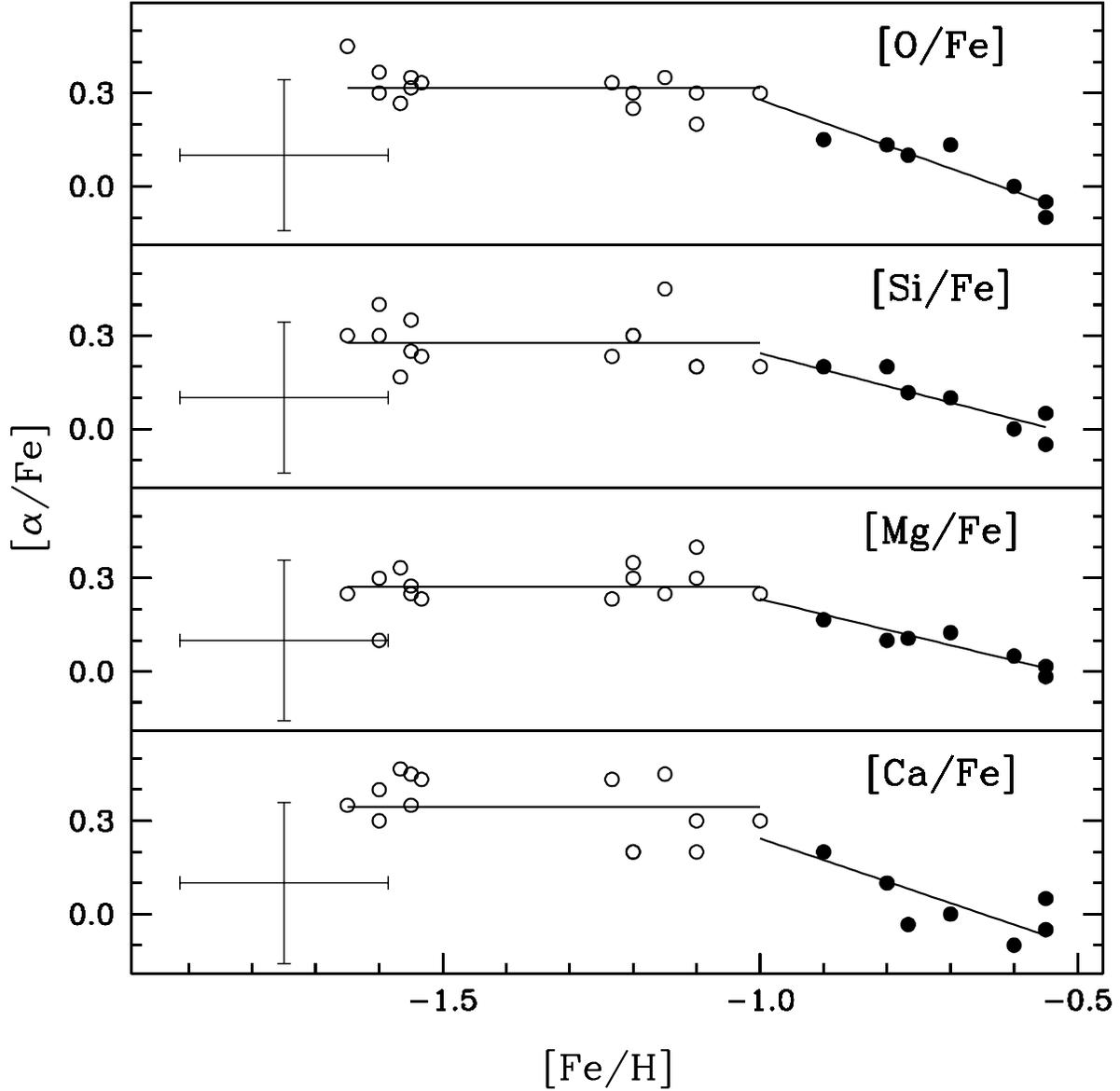}      
\caption{
[Ca/Fe], [Mg/Fe], [Si/Fe] and [O/Fe] abundance ratios as a function of 
[Fe/H] for the RGB stars in our sample (cf. Sect.~4 and Table~1).  
Open circles refer to the {\it normal} metal-poor and intermediate 
metallicity populations, 
filled circles to the {\it anomalous}, metal rich one.
The continuous lines are the least square fitting to the data, 
while the crosses on the bottom-left corner represent the average 
error (over the full sample of stars) on the abundance ratio estimates.}
\end{figure}  

\begin{figure}
\epsscale{1.0}
\plotone{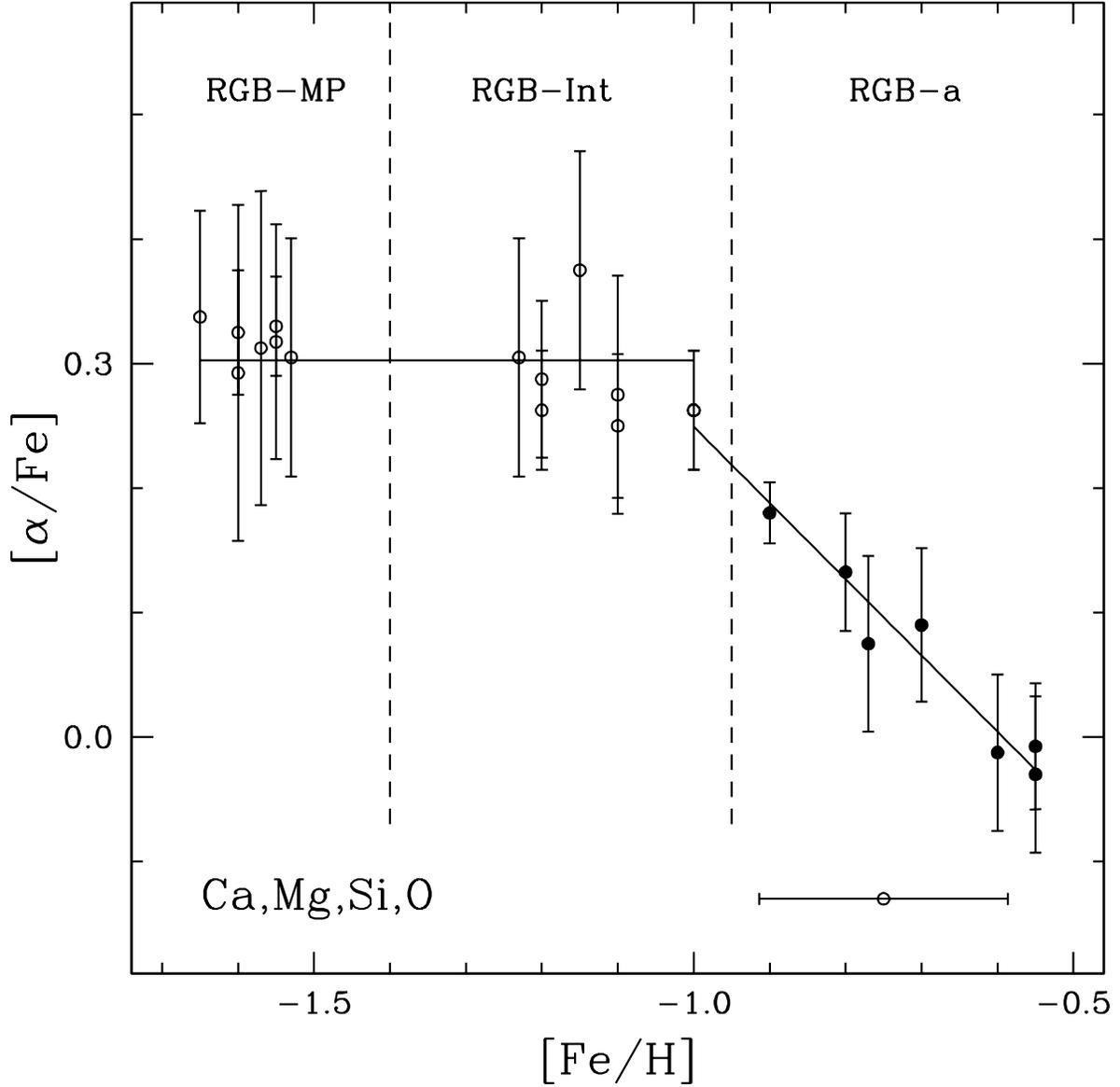}      
\caption{
[$\alpha $/Fe] abundance ratios 
as a function of [Fe/H]
for the observed RGB stars, as inferred by averaging the 
[Ca/Fe], [Mg/Fe], [Si/Fe] and [O/Fe] relative abundances 
(cf. Sect.~4 and Table~1).  
Symbols as in Fig.~5.
The continuous lines are the least square fitting to the data, 
the error bar on the bottom-right corner is the average error 
(over the full sample of stars) on the [Fe/H] abundance  
determination, while the error bars along the y-axis refer to the  
error on the average [$\alpha$/Fe] estimate for each star.}
\end{figure}  

\begin{figure}
\epsscale{1.0}
\plotone{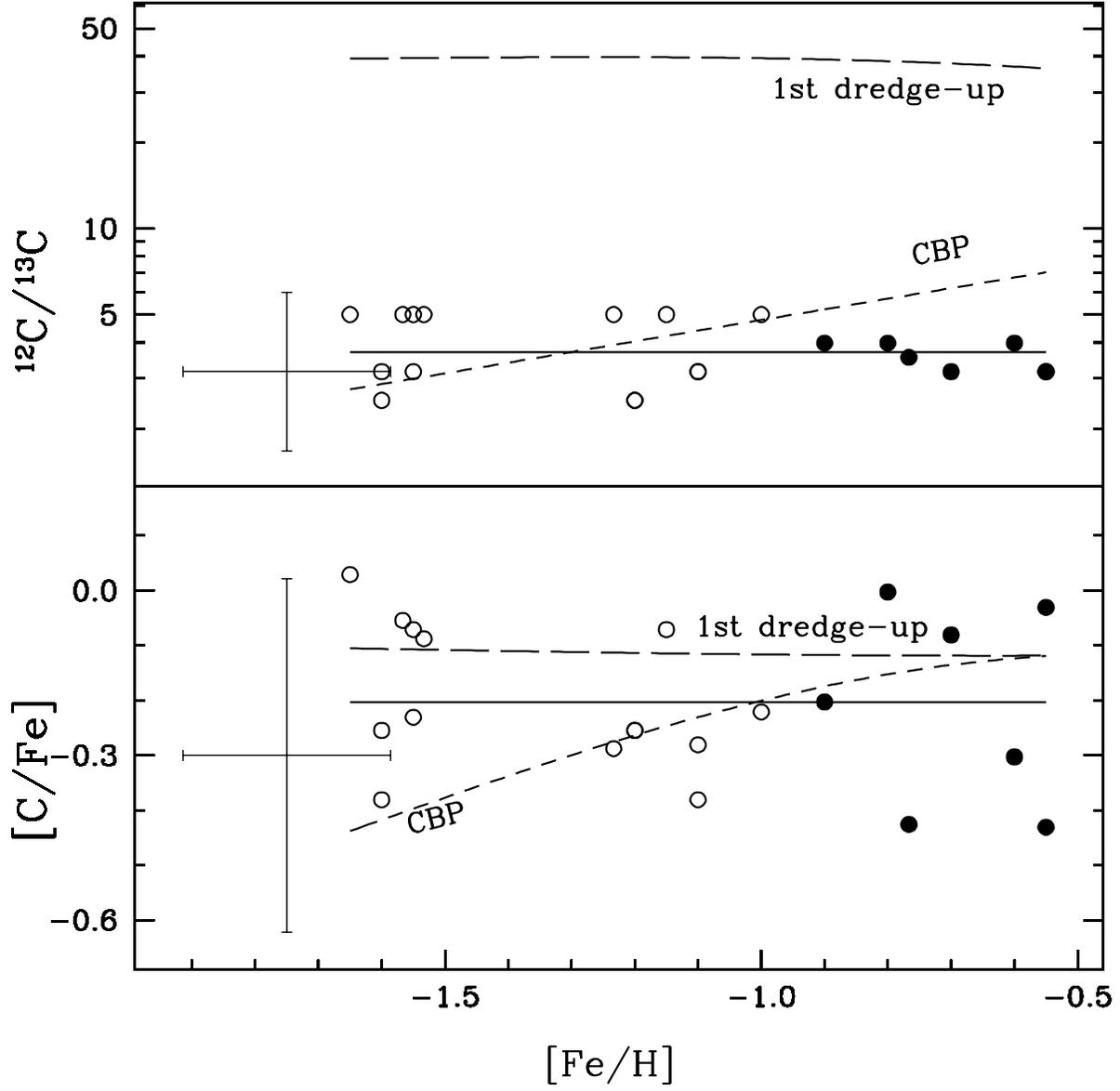}      
\caption{
[C/Fe] (bottom panel) and $^{13}$C/$^{12}$C (top panel) abundance ratios 
as a function of [Fe/H] for the observed RGB stars
(cf. Sect.~4 and Table~1).  
Symbols as in Fig.~5.
The continuous lines are the least square fitting to the data, 
while the long-dashed and short-dashed lines refer to 
the {\it first dredge-up} and {\it cool bottom process} modeling by 
Boothroyd \& Sackmann (1999), respectively.
The crosses on the bottom-left corner represent the average 
error (over the full sample of stars) on the abundance ratio estimates.}
\end{figure}  

\end{document}